\documentclass[aps,twocolumn,preprintnumbers,nofootinbib,superscriptaddress]{revtex4}
\usepackage{times}
\usepackage{amsmath}
\usepackage{amssymb}
\usepackage{graphicx}
\usepackage{subfigure,accents}
\usepackage{booktabs}
\usepackage{rotating}
\usepackage{longtable}
\usepackage{bm}
\usepackage[pagebackref=false,colorlinks=,linkcolor=blue,citecolor=blue]{hyperref}
\usepackage{epstopdf}
 \pdfoutput=1
 \usepackage[utf8]{inputenc}
\usepackage{cases}
\usepackage{amsmath,float}
\usepackage{amssymb}
\usepackage{amsfonts}
\usepackage{amssymb}
\usepackage{dcolumn}
\usepackage{bm}
\usepackage{bbm}
\usepackage{graphicx}
\usepackage{xcolor}
\usepackage{array}
\usepackage{subfigure}
\usepackage{hyperref}
\usepackage{wasysym}

\newcommand{\be}{\begin{equation}}
\newcommand{\ee}{\end{equation}}
\newcommand{\ba}{\begin{eqnarray}}
\newcommand{\ea}{\end{eqnarray}}

\newcommand{\gsim}{\mathrel{\hbox{\rlap{\lower.55ex \hbox {$\sim$}}
                   \kern-.3em \raise.4ex \hbox{$>$}}}}
\newcommand{\lsim}{\mathrel{\hbox{\rlap{\lower.55ex \hbox {$\sim$}}
                   \kern-.3em \raise.4ex \hbox{$<$}}}}

\hypersetup{colorlinks=true,
            breaklinks=true,
            pdfstartview=Fit,
            linkcolor=blue,
            citecolor=blue,
            urlcolor=blue}
            
 \usepackage{graphicx}

\usepackage{amsmath}

\usepackage{braket}

\newcommand{\bw}{\begin{widetext}}
\newcommand{\ew}{\end{widetext}}

\usepackage{mathtools}

\def\ber{\begin{eqnarray}}
\def\eer{\end{eqnarray}}
\def\beq{\begin{equation}}
\def\eeq{\end{equation}}

\begin{document}

\title{Black holes surrounded by Einstein clusters as models of dark matter fluid}

\author{Kimet Jusufi} \email{kimet.jusufi@unite.edu.mk}
\affiliation{Physics Department, State University of Tetovo, Ilinden Street nn, 1200, Tetovo, North Macedonia}

\begin{abstract}
    We construct a novel class of spherically symmetric and asymptotically flat black holes and naked singularities surrounded by anisotropic dark matter fluid with the equation of state (EoS) of the form $P_t=\omega \rho$.  We assume that dark matter is made of weakly interacting particles orbiting around the supermassive black hole in the galactic center and the dark matter halo is formed by means of Einstein clusters having only tangential pressure. In the large distance from the black hole we obtain the constant flat curve with the upper bound for the dark matter state parameter $\omega\lsim 10^{-7}$. To this end, we also check the energy conditions of the dark matter fluid outside the black hole/naked singularity, the deflection of light by the galaxy, and the shadow images of the Sgr A$^\star$ black hole using the rotating and radiating particles. For the black hole case, we find that the effect of dark matter fluid on the shadow radius is small, in particular the angular radius of the black hole is shown to increase by the order $10^{-4} \mu$arcsec compared to the vacuum solution. For the naked singularity we obtain significantly smaller shadow radius compared to the black hole case. Finally, we study the stability of the S2 star orbit around Sgr A$^\star$ black hole under dark matter effects. It is argued that the motion of S2 star orbit is stable for values $\omega \lsim 10^{-7}$, however further increase of $\omega $ leads to unstable orbits.  Using the observational result for the shadow images of the Sgr A$^\star$ reported by the EHT along with the tightest constraint for $\omega$ found from the constant flat curve, we show that the black hole model is consistent with the data while the naked singularity in our model can be ruled out. 
\end{abstract}
\maketitle

\section{Introduction}
Black holes are extremely interesting objects predicted to exists by general relativity. Through many astrophysical observations, including the recent observation of the image of M87 and Sgr A$^\star$ supermassive black holes by Event Horizon Telescope (EHT) collaboration \cite{m87,m871,EHT2022-1,EHT2022-2,EHT2022-3}, detection of X-rays and gravity waves \cite{1,2,3}, in fact it is now widely believed that in the center of each galaxy there are objects described just by the spacetime geometry of black holes predicted by Einstein's theory.  In addition, black holes are now considered as astrophysical laboratories for testing not only general relativity in the strong gravity regime but also different extensions or modified theories of gravity.  On other hand,  black holes have played an important role in discovering a deep connection between the laws of thermodynamics and gravity, this can play a significant role for understanding the quantum nature of black holes. We now know that black holes have entropy and as was shown by Hawking they radiate energy as a consequence an external observer located far away from the black hole should detect temperature \cite{4,5}. In that sense, black holes are important for testing quantum theories of gravity as well. 

In spite of the great success of general relativity, there are many open problems that general relativity cannot solve. One can mention here the problem with the black hole singularities, cosmic inflation, constant flat curves in galaxies \cite{6}, the accelerated expansion of the universe, and so on. The problem with the rotating curves in galaxies can be explained with the dark matter - a mysterious substance probably made of a new particle which weakly interacts with the surrounding  and hence very difficult to be detected. We can mainly probe the dark matter effect in terms of the gravity.  An alternative way to explain the constant flat curves is to modify the law of gravity \cite{7}, but other interesting possibilities have been suggested recently, like dark matter as Bose-Einstein condensate \cite{8}, dark matter as a superfluid \cite{9}, emergent gravity \cite{10} and others.  As of today, there is no definite answer and this questions and the problem of dark matter remains an open problem in physics.

In this work, we shall focus on an old method developed by Einstein to construct spherically symmetric solutions using the anisotropic matter distributions known as the Einstein clusters \cite{11}. Moreover, this idea was used to explain dark matter in terms of Einstein clusters by Boehmer and Harko \cite{12} and Lake \cite{13}, as well as other related studies \cite{R1,R2,R3}. Recently Cardoso et al. \cite{14} used the Einstein cluster to construct a space time geometry using the Hernquist-type distribution and a black hole in the center. In this work, we are interested to model the dark matter as a fluid with EoS given by $P_t=\omega \rho$ along with a central supermassive black hole. We shall consider a different mass profile which locally diverges but globally reduces to a asymptotically flat spacetime. Moreover we would like to understand more about the effect of dark matter on shadow images of the Sgr A$^\star$ black hole, it's effect on the motion of S2 star in our galactic center, and the constraint on the dark matter parameter $\omega$. Toward this goal, we are going to extend the non-minimally coupling effect between the black hole geometry and the dark matter suggested in \cite{14} and explore in details two special solutions. 

The paper is organized as follows. In Section 2, we briefly review the construction by Einstein and then we introduce a black hole and a naked singularity in a dark matter fluid. In Section 3, we explore the deflection of light by the galaxy due to the dark matter and we also discuss the constant flat curve and the constrain on $\omega$. In Section 4, we study the shadow images using rotating and radiating particles near the black hole and a naked singularity, respectively.  Importantly, we will use the EHT result for the Sgr A$^\star$ shadow radius to test our models. In Section 5, we study the S2 star orbit. We comment on our results in the last section.

\section{A black hole surrounded by anisotropic dark matter fluid}
Let us start by assuming a surrounding anisotropic matter which according to the Einstein construction can be written in terms of the average stress tensor given by  \cite{11,12,13,14}
\begin{equation}
\langle T^{\mu\nu}\rangle=\frac{n}{m_p}\langle P^{\mu}P^{\nu}\rangle,
\end{equation}
where $n$ is the number density of particles, $m_p$ is the mass of the particle and $P^{\mu}$ is the four-momentum satisfying the geodesic equations. According to this model, we have anisotropic  matter with only tangential pressure $P_t$ and vanishing radial pressure. In particular we can write   \cite{11,12,13,14}
\be
T^\mu_\nu={\rm diag}(-\rho,0,P_t,P_t)\,.
\ee 

In this work we are interested to describe a spherically symmetric geometry therefore we can use the following line element 
\be
ds^2=-f(r)dt^2+\frac{dr^2}{g(r)}+r^2(d\theta^2+\sin^2\theta d\phi^2),
\label{eq:Spherical}
\ee
with 
\begin{equation}
    g(r)=1-\frac{2m(r)}{r}.
\end{equation}

One can show that the radial function $f(r)$ is related to the mass function according to the equation  \cite{14}
\be
\frac{rf'(r)}{2f(r)}=\frac{m(r)}{r-2m(r)}.\label{eq_metric}
\ee
On the other hand, one can use the conservation of the energy, to show a relation between the tangential pressure and the mass profile given by \cite{14}
\be
\frac{2P_t}{\rho}=\frac{m(r)}{r-2m(r)}.
\ee
In this paper we assume that dark matter is made of particles rotating around the black hole and the resulting dark matter halo can be expressed as an effective fluid with the EoS of the form $P_t=\omega \rho$. From the last equation, it is easy to show that the mass profile reads 
\begin{equation}
m(r)= \frac{2 \omega r}{1+4 \omega}.
\end{equation}
Moreover from the last equation we can now obtain the energy density of the dark matter fluid
\begin{eqnarray}
\rho=\frac{1}{4 \pi r^2} \frac{dm(r)}{dr}=\frac{\omega}{2 \pi r^2 (1+4 \omega)}.
\end{eqnarray}

It is very interesting to see that our model of Einstein cluster is analogues to the isothermal dark matter profile which can be written as 
\begin{equation}
\rho(r)=\frac{\mathcal{C}}{4 \pi r^2},
\end{equation}
where we have defined 
\begin{equation}
\mathcal{C}=\frac{2 \omega}{1+4 \omega}.
\end{equation}

Later on, we shall assign a physical interpretation to this quantity. Furthermore, for the tangential pressure we obtain 
\begin{eqnarray}
P_t=\frac{\omega^2}{2 \pi r^2 (1+4 \omega)}.
\end{eqnarray}

Let us now introduce a black hole in our spacetime in the spirit of Ref. \cite{14}. In order to include the central object in the center, we can modify the mass profile solution by including a coupling between the central object (for example a black hole) and dark matter, hence by generalizing the expression for the mass profile as follows
\begin{equation}
m(r)= M_{0}+\frac{2 \omega r}{1+4 \omega}\left(1-\frac{2M_{0}}{r}\right)^{\alpha}.
\end{equation}
In general, we see that such a mass profile would lead to a divergent mass and hence would be incompatible with asymptotically flat spacetime, however, our goal here is to construct asymptotically flat spacetimes from this mass profile. Note also that $\alpha$ is some real number, in what follows we shall consider some special cases. As we are going to see, by assuming a central object with mass $M_0$, in general, depending on the specific choice of the mass function one can end up not only with a black hole but also with a naked singularity. 

\subsection{Model 1}
In our first example,  we are going to consider the simplest case $\alpha=1$ in the mass profile equation. Our goal is to find the radial function, hence we can use Eq.(5) to obtain 
\begin{eqnarray}
(r^2-2 M_{0} r)f'(r)-2f(r)(2 r \omega+M_{0})=0.
\end{eqnarray}

Interestingly, as we shall see the dark matter profile has a very simple solution for the metric components. Solving this equations and using the condition $f(r)|_{r=R}=1$ to fix the constant of integration and find the solution 
\begin{eqnarray}
f(r)= \frac{\left(1-\frac{2M_{0}}{r}\right)}{\left(1-\frac{2M_{0}}{R}\right)} \left(\frac{r-2 M_{0}}{R-2 M_{0}} \right)^{4 \omega}. 
\end{eqnarray}

At this point we can rescale the time coordinate and absorb the factor $1-2M_{0}/R$ as follows 
\begin{eqnarray}
dt' \to \frac{dt}{\sqrt{1-\frac{2M_{0}}{R}}}.
\end{eqnarray}
Therefore, we can say that our solution in the general case describes the whole region with the following spacetime metric [by dropping the prime notation in the time coordinate] 
\begin{widetext}
\[  ds^2=
\begin{dcases} 
     -\left(1-\frac{2M_{0}}{r}\right) \left(\frac{r-2 M_{0}}{R-2 M_{0}} \right)^{4 \omega} dt^{2}+\frac{dr^{2}}{\left(1-\frac{2M_{0}}{r}\right)\left(1-\frac{4 \omega}{1+4 \omega} \right)}+ r^{2} d\Omega^{2},  & r< R \\
      -\left(1-\frac{2M}{r}\right) dt^{2}+\frac{dr^{2}}{1-\frac{2 M}{r}}+r^{2} d\Omega^{2}, & r\geq R
   \end{dcases}
   \] 
\end{widetext}
Basically, we have the interior region of the galaxy described by the line element in the region $r<R$, and the exterior region $r\geq R$ which is effectively a Schwarzschild metric.  Note that as a special case by setting $\omega=0$ we can obtain the vacuum solution without dark matter effect.  Furthermore it looks like this solution has an event horizon at $r=2 M_{0}$, but a further investigation shows that the metric inside $r<R$ is not a black hole, rather it is a naked singularity. One way to see this is by looking at the regularity condition at $r=2M_{0}$, and by computing $\det g_{\mu\nu}$ evaluated at $r=2M_{0}$ which gives zero.  In other words, the central singularity has been shifted at $r=2M_{0}$. We can see this more clearly if we introduce new coordinates $r_{new}\to r-2M_0$ and $R \to R-2M_0$, where $M_{0}$ is now the mas parameter of the naked singularity. The interior metric in terms of the new coordinate reads 
\begin{widetext}
\[  ds^2=
\begin{dcases} 
     -\left(\frac{r}{r+2M_0}\right) \left(\frac{r}{R} \right)^{4 \omega} dt^{2}+\frac{dr^{2}}{\left(\frac{r}{r+2M_0}\right)\left(1-\frac{4 \omega}{1+4 \omega} \right)}+ (r+2M_0)^{2} d\Omega^{2},  & r< R \\
      -\left(1-\frac{2M}{r}\right) dt^{2}+\frac{dr^{2}}{1-\frac{2 M}{r}}+r^{2} d\Omega^{2}. & r\geq R
   \end{dcases}
   \] 
\end{widetext}
The interior solution as can be seen is regular at $r=2M_0$, but is has a central singularity at $r=0$, thus it is a naked singularity. The region $r>R$ remains basically the same in large distances from the center. The mass of the system is given by
\begin{equation}
M= M_{0}+\frac{2 \omega R}{1+4 \omega}\left(1-\frac{2M_{0}}{R}\right).
\end{equation}

In the last equation $R$ is some large number and in principle can be obtained from observations if we know the total mass and the parameter $\omega$. Later on, we will discuss this issue in more details. 
As we already pointed out the mass profile in Eq. (12) diverges, in order to obtain asymptotically flat solution in the large limit of $r$ the whole process must be conducted with caution.  Namely, we can see that the interior metric solves the Einstein equations while the exterior the vacuum solution, in addition to that the metric is continuous but not smooth at $r=R$. Such a similar issue was recently discussed by Remmen (see for more details \cite{remmen}) in the context of singular isothermal fluid.  
To resolve this issue, at the surface at $\Sigma$, we need to discuss the junction conditions.  In other words, at the surface $\Sigma$, where the dark matter fluid in the interior region matches the effective Schwarzschild vacuum exterior, there must be a thin shell of matter that confines the fluid. The energy-momentum for the induced metric on $\Sigma$ reads $T_{ab} = -\sigma (u_a u_b + \gamma_{ab} )$, where $\sigma$ bis the surface density energy.  In the full spacetime coordinates, we can write the extra energy-momentum tensor of the shell in the following form \cite{remmen}
\be 
\Delta T_{\mu\nu} = -\sigma\,\delta(r-R)\, {\rm diag}(0,0,r^2,r^2\sin^2\theta)|_{r=R}.\label{eq:surface}
\ee

As we are going to see the energy-momentum tensor of the surface violates the null energy condition, but as pointed out in \cite{remmen} this can be simply rectified by adding an arbitrary surface density larger than $\sigma$. We can compute the $a^\mu = (u^\nu \nabla_\nu u^\mu)$, that is, the acceleration in both regions resulting with 
\be 
a =  
\begin{dcases}
\frac{4\omega r+4 \omega M_0+M_{0}}{2 r (r+M_0) \sqrt{1+4 \omega} }\left(1+\frac{2M_{0}}{r}\right)^{-1/2} & r<R \\
\frac{M}{r^2} \left(1-\frac{2M}{r}\right)^{-1/2} & r \geq R.
\end{dcases}
\ee
As we know the surface gravity of the object is the force at infinity needed to suspend an observer at the $r=R$ surface. At the distance $r=R$, since $M_{0}<<R$, thus we can basically neglect the central mass and set it to zero, i.e., $M_{0}\to 0$, and simplify the computations. In that case, taking the limit from the inside versus outside we indeed find a discontinuity
\be
\begin{aligned}
\kappa_+ &= \lim_{r\rightarrow R^+} |\xi| a = \frac{M}{R^2},\\
\kappa_- &= \lim_{r\rightarrow R^-} |\xi|a =   \sqrt{1+4 \omega} \frac{M}{R^2 },
\end{aligned} 
\ee
note that one can use the Killing vector $\xi^{\mu}$ to find $|\xi|$. We now see that the difference across the $r=R$ boundary gives
\be
\Delta \kappa = \kappa_+ - \kappa_- = \frac{2 \omega }{R}\left(  \frac{1-\sqrt{1+4 \omega}}{1+4 \omega}\right), 
\ee
using the expansion 
\begin{eqnarray}
\sqrt{1+4 \omega}=1+2 \omega+...
\end{eqnarray}
we can approximate the difference as 
\be
\Delta \kappa \simeq  -\frac{2\,\mathcal{C} \omega }{R} = - 8\pi \sigma, 
\ee
where $\mathcal{C}$ is defined by Eq. (10). From the last equation we see that the corresponding perfect fluid to the thin shell has surface tension given by $\sigma =  \mathcal{C} \omega/4\pi R$. One can relate this to the the extra force needed to counteract the tension supplied by the boundary shell. The mass of the thin shell can be computed from the relation 
\begin{eqnarray}
\frac{\partial M_{\rm shell}}{\partial A}=\frac{\Delta \kappa}{8 \pi}=\sigma,
\end{eqnarray}

This means that the only effect on the geometry of doing so would be to shift $M$ in the $r>R$ part of the metric by the total mass of the shell, which gives 
\begin{eqnarray}
M_{\rm tot}=M+M_{\rm shell}=\frac{2 \omega R}{1+4 \omega}\left(1-\frac{\omega}{1+4 \omega}\right).
\end{eqnarray}
We found that the corrections due to the thin shell matter are proportional to $\omega^2$, but as  we are going to see $\omega$ is a very small number, hence the total mass of the dark matter halo can be approximated to $M_{\rm tot}\simeq M$ and, more importantly, we will have an asymptotically flat spacetime geometry. In Fig. 1, we show the plot of the total mass as a function of the state parameter $\omega$ and $R$. 

\subsubsection{Energy conditions}
It is interesting to elaborate the energy conditions for the dark matter fluid. They are a sets of inequalities depending on energy momentum tensor. In particular the weak energy condition (WEC), i.e. $T_{\mu\nu} U^{\mu}U^{\nu}$, where $U^{\mu}$ is a timelike vector implies
\begin{equation}\label{EC1}
\rho (r)\geq 0 ~~\text{and}~~ \rho (r)+P_{i}(r)\geq 0.
\end{equation}
On the other hand, the null energy condition (NEC) is given by $T_{\mu\nu} k^{\mu} k^{\nu}$, where $k^{\mu}$ is null vector. This implies that $\rho (r)+P_{i}(r)\geq 0$, with $i= 1,2,3$. 

We also have the strong energy condition (SEC) given by,
\begin{equation}\label{EC3}
\rho (r)+ \sum P_{i}(r)\geq 0, ~~\text{and}~~ \rho (r)+P_{i}(r)\geq 0.
\end{equation}

From the energy density of the dark matter fluid we have 
\begin{equation}
    \rho (r)+P_{r}(r)|_{r=r_0}=\frac{\omega}{2 \pi r_0^2 (1+4 \omega)}\geq 0,
\end{equation}
where in our case we have $P_r(r)=0$. Since $r_0^2>0$, from the last equation it follows that $\omega \geq 0$. Furthermore we can check the strong energy condition to find
\begin{equation}
\rho (r)+P_t(r)|_{r=r_0}=\frac{\omega(1+\omega)}{2 \pi r_0^2 (1+4 \omega)}\geq 0.
\end{equation}

Here the positive region is given in the interval $-1 \leq \omega < -1/4$ and $\omega \geq 0$, therefore we choose as a physical solution only the region $\omega \geq 0$.
Finally using Eq. (\ref{EC3}), we obtain 
\begin{equation}
\rho (r)+P_{r}(r)+2P_t(r) =\frac{\omega(1+2\omega)}{2 \pi r_0^2 (1+4 \omega)}\geq 0.
\end{equation}
the positive region is given in the interval $-1/2 \leq \omega < -1/4$ and $\omega \geq 0$, again we choose only the region $\omega \geq 0$.\\

Thus, we argued that the energy conditions in general are satisfied by the dark matter fluid outside the naked singularity metric. In what follows, we are going to consider another important mass profile which results with a black hole in the center.   

\begin{figure}[!htb]
 	\includegraphics[width=8.5 cm]{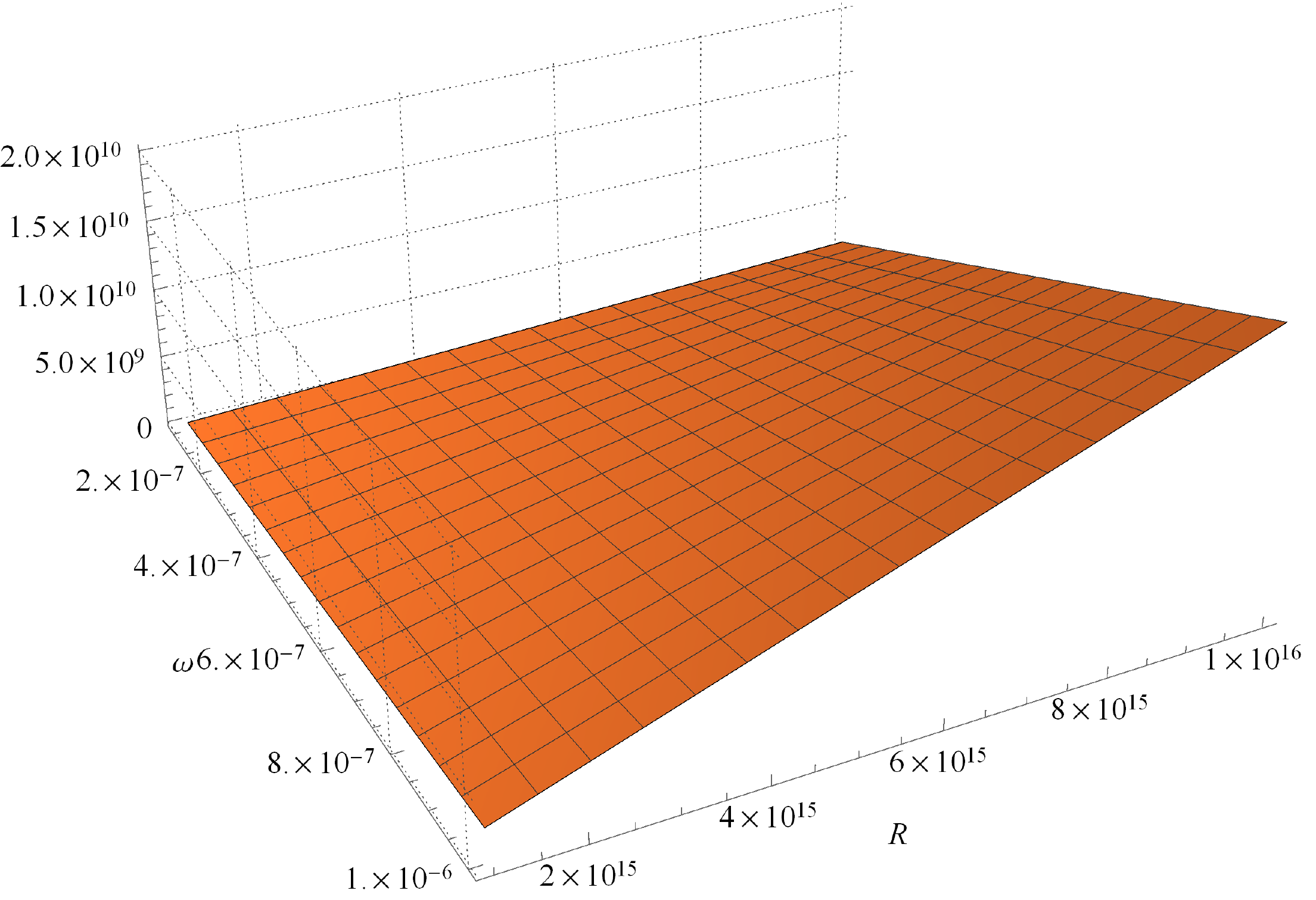}
		\caption{{The plot shows the total mass of the dark matter halo as a function of state parameter $\omega$ and $R$. 
 }}
	\end{figure}

\subsection{Model 2}
Here we consider another solution with a slightly different and more realistic mass profile for the dark matter fluid. As we shall see, in this case we do get a black hole metric, hence we replace $M_0=M_{BH}$ in the mass profile. Specifically, we will focus and explore here the case $\alpha=2$ in our mas profile.  From Eq.(5) we find the following differential equation
\begin{eqnarray}\notag
&&\frac{\left(9r-2 M_{BH}\right)\left(r+8 \omega M_{BH}\right)}{1+4 \omega}f'(r)\\
&-& 2f(r)\left[M_{BH}+\frac{2 \omega r}{1+4\omega} \left(1+\frac{2M_{BH}}{r}\right)^2 \right]=0.
\end{eqnarray}

By solving this differential equation and by using the condition $f(r)|_{r=R}=1$ and after we fix the constant of integration and find the solution 
\begin{eqnarray}
f(r)= \frac{\left(1-\frac{2M_{BH}}{r}\right)}{\left(1-\frac{2M_{BH}}{R}\right)} \left(\frac{r+8 \omega M_{BH}}{R+8 \omega M_{BH}} \right)^{4 \omega}. 
\end{eqnarray}

In a similar way,  we can absorb the factor $1-2M_{BH}/R$ in the time coordinate, in that case it follows that our solution has the following spacetime metric in two regions 
\begin{widetext}
\[  ds^2=
\begin{dcases} 
     -\left(1-\frac{2M_{BH}}{r}\right)\left(\frac{r+8 \,\omega\, M_{BH}}{R+8\, \omega\, M_{BH}} \right)^{4 \omega} dt^{2}+ \frac{dr^{2}}{\left(1-\frac{2M_{BH}}{r}\right)\left[1-\frac{4 \omega}{1+4 \omega} \left(1-\frac{2M_{BH}}{r}\right)\right]}+ r^{2} d\Omega^{2}.  & r< R \\
      -\left(1-\frac{2M}{r}\right) dt^{2}+\frac{dr^{2}}{1-\frac{2 M}{r}}+r^{2} d\Omega^{2} & r\geq R
   \end{dcases}
   \] 
\end{widetext}
where 
\begin{equation}
M= M_{\rm BH}+\frac{2 \omega R}{1+4 \omega}\left(1-\frac{2M_{\rm BH}}{R}\right)^2.
\end{equation}

In this case, the metric has an event horizon at $r=2M_{BH}$ and the regularity condition at $r=2M_{BH}$ is satisfied. One can simply check that $\det g_{\mu\nu}$ evaluated at $r=2M_{BH}$ is a positive number.  
At this point, we need to discuss the junction conditions at $r=R$. Following the same approach as in the case of Model 1, therefore we can skip the details here and simply compute the acceleration in both regions 
\be 
a =  
\begin{dcases}
\frac{2 r^2 \omega  +(1-4 \omega) M_{BH}r +8 M_{BH}^2 \omega }{r^2 \sqrt{1+4 \omega} \sqrt{(r-2M_{BH})(r+8 \omega M_{BH})} }, & r<R \\
\frac{M}{r^2} \left(1-\frac{2M}{r}\right)^{-1/2}, & r \geq R.
\end{dcases}
\ee

By taking the limit from the inside versus outside we again obtain a discontinuity according to 
\be
\begin{aligned}
\kappa_+ &= \lim_{r\rightarrow R^+} |\xi| a = \frac{M}{R^2}\\
\kappa_- &= \lim_{r\rightarrow R^-} |\xi|a =   \sqrt{1+4 \omega} \frac{M}{R^2 }.
\end{aligned} 
\ee

Where we have simplified the computations by neglecting the black hole mass. It follows that we obtain the same difference across the $r=R$ boundary as in the Model 1, that is $\Delta \kappa =-2 \mathcal{C}\omega/R$. Using the relation between the surface density and the mass of the shell we find that the total mass of the system is given by Eq. (24). Again, the corrections due to the thin shell of matter are very small and the resulting spacetime geometry is asymptotically flat. 
	
\subsubsection{Energy conditions}
Finally, we can elaborate the energy conditions using the dark matter Model 2. From the energy density it is easy to show 
\begin{equation}
   \rho (r)+P_{r}(r)|_{r_0}= \frac{\omega (r_0^2-4M_{BH}^2)}{2 \pi r_0^4 (1+4 \omega)}\geq 0.
\end{equation}
Since $r_0^4>0$, it follows that outside the black hole we must have  $r_0>2M_{BH}$, and hence one must have $\omega \geq 0$.  On the other hand, the (SEC) stipulates that
\begin{equation}
\rho (r)+P_t(r)|_{r_0}=\frac{\omega (1+\omega) (r_0^2-4M_{BH}^2)}{2 \pi r_0^4 (1+4 \omega)} \geq 0.
\end{equation}

In this case, we again can see that outside the event horizon the positive interval is given by $-1 \leq \omega < -1/4$ and $\omega \geq 0$, but we accept as a physical solution only the region $\omega \geq 0$.
We easily find also that 
\begin{equation}
\rho (r)+P_{r}(r)+2P_t(r)|_{r_0}=\frac{\omega (1+2 \omega) (r_0^2-4M_{BH}^2)}{2 \pi r_0^4 (1+4 \omega)}  \geq 0,
\end{equation}
which has a positive region outside the black hole when  $-1/2 \leq \omega < -1/4$ and $\omega \geq 0$. This shows that, the energy conditions are indeed satisfied only in the domain $\omega \geq 0$. Thus, we have shown that the energy conditions are satisfied by the dark matter fluid outside the black hole having $\omega>0$. The value of $\omega$ can only be obtained from observations.

\section{The constant flat curve, deflection of light by a galaxy, and the  Einstein rings}
Near the region $r\gsim R$, the constant flat curve for the velocity should be recovered. The solution in this region as we pointed out is asymptotically flat and given by 
\begin{equation}
   ds^2=  -\left(1-\frac{2M_{\rm tot}}{r}\right) dt^{2}+\frac{dr^{2}}{1-\frac{2 M_{\rm tot}}{r}}+r^{2} d\Omega^{2},
\end{equation}

In the limit $r \to \infty$, we can neglect the black hole mass thus we end up 
\begin{equation}
M_{\rm tot} \simeq \frac{2 \omega R}{1+4 \omega}\left(1-\frac{\omega}{1+4 \omega}\right),
\end{equation}
and the spacetime element reads 
\begin{eqnarray}
 ds^{2}&=& \notag
-\left[1-\frac{4 \omega R}{r(1+4 \omega)} \left(1-\frac{\omega}{1+4 \omega}\right) \right] dt^{2}\\
&+& \frac{dr^{2}}{\left[1-\frac{4 \omega R}{r(1+4 \omega)} \left(1-\frac{\omega}{1+4 \omega}\right) \right]}+r^{2} d\Omega^{2}.
\end{eqnarray}

This metric is suitable to study the deflection of light by a galaxy in presence of dark matter. In leading order terms the deflection angle of light by the galaxy easily can be found to be
\begin{eqnarray}
\hat{\alpha}=\frac{8 \omega R}{b (1+4 \omega)} \left(1-\frac{\omega}{1+4 \omega}\right).
\end{eqnarray}

We can specialize the deflection angle when $b \sim R$, in that case we  we see that the deflection angle will be independent of the impact parameter
\begin{eqnarray}
\hat{\alpha}=\frac{8 \omega }{1+4 \omega} \left(1-\frac{\omega}{1+4 \omega}\right).
\end{eqnarray}

This is very similar to the isothermal dark matter profile which has the deflection angle $\hat{\alpha}_{IS}=2 \pi v_{0}^2$. This equation can explain the flat rotation curve in the outer region of the galaxy for large $r>>M_{BH}$ in the region $r \sim R$. Namely, we can neglect the term $\omega^2$ and the black hole mass for such large distances, then the velocity goes like 
\begin{eqnarray}
v_{0}^2 \sim \frac{2 \omega}{1+4 \omega}.
\end{eqnarray}
In terms of the velocity, our result can be written as $\hat{\alpha}=4 v_{0}^2$, which is slightly smaller compared to the isothermal dark matter model. If we take as an example $v_{0} \sim 220$km/s, we find $\omega \sim 2.68 \times 10^{-7}$, and a deflection angle $\hat{\alpha} \sim 2.15 \times 10^{-6}$. Compared to the case of isothermal dark matter profile $\hat{\alpha}_{IS} \sim 6.28 \times 10^{-6}$, this suggest that a galactic dark matter halo modelled as Einstein cluster predicts slightly smaller gravitational lensing effect.  A similar effect was found in \cite{12}. In Eq. (39) if we assume a dark matter mass $M_{tot} \sim 10^{10} M_{BH}$ in our galaxy, we find $R \sim 1.85 \times 10^{16} M_{BH}$. 

The small angles lens equation (in the weak deflection approximation) reads
\begin{eqnarray}\label{EinsteinRing}
    \beta=\theta-\frac{D_{LS}}{D_{OS}}\hat{\alpha}.
\end{eqnarray}
In the special situation $\beta=0$, when the source lies on (or passes through) the optical axis an Einstein ring is formed. The weak deflection approximation, $\hat{\alpha} \ll 1$ represents the angular radius of the Einstein ring given by
\begin{eqnarray}\label{EinsteinRing1}
    \theta_{E}\simeq\frac{D_{LS}}{D_{OS}}\hat{\alpha}(b).
\end{eqnarray}
Here we took into account that $D_{OS}=D_{OL}+D_{LS}$, when the angular source position is $\beta=0$. Keeping only the first order of the deflection angle and using the relation $b=D_{OL}\sin{\theta}\simeq D_{OL}\theta$ the bending angle in the small angle approximation gives the the angular radius
\begin{eqnarray}
    \theta_{E}\simeq \sqrt{\frac{D_{LS}}{D_{OS}D_{OL}}\left[\frac{8 \omega R}{1+4 \omega} \left(1-\frac{\omega}{1+4 \omega}\right)\right]}.
\end{eqnarray}
	
The lensing phenomena is a great tool to test different dark matter models based on the prediction of the deflection angle. Although the difference is small, in principle, we can distinguish different models.

\begin{figure*}[!htb]
 	\includegraphics[width=8.9 cm]{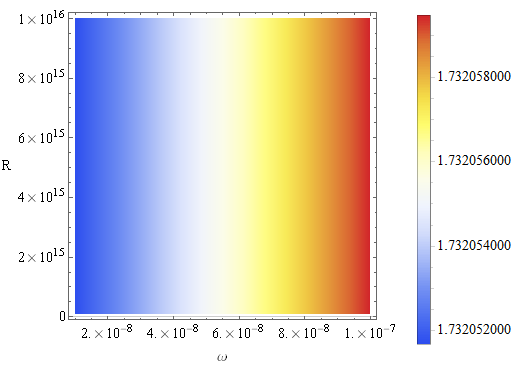}
 		\includegraphics[width=8.9 cm]{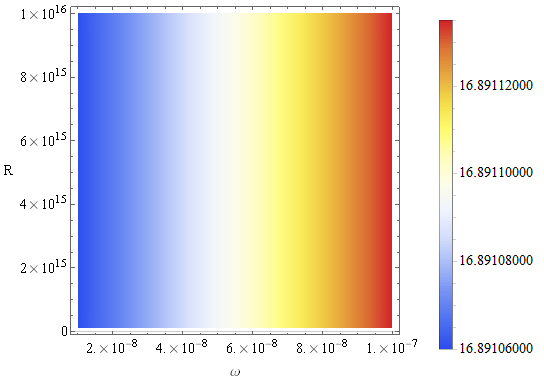}
		\caption{{\it{Shadow radius (left panel) and angular diameter (right panel) of the Sgr A$^\star$ using the Model 1. We have set the location of the observer $r_0=10^{10}$ and $M_{0}=4.1 \times 10^6 M_{\odot}$.
 }}}
\label{Densityplot}
	\end{figure*}
	
\begin{figure*}[!htb]
 	\includegraphics[width=8.9 cm]{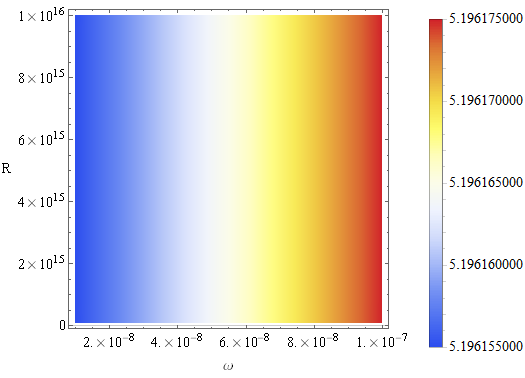}
 		\includegraphics[width=8.9 cm]{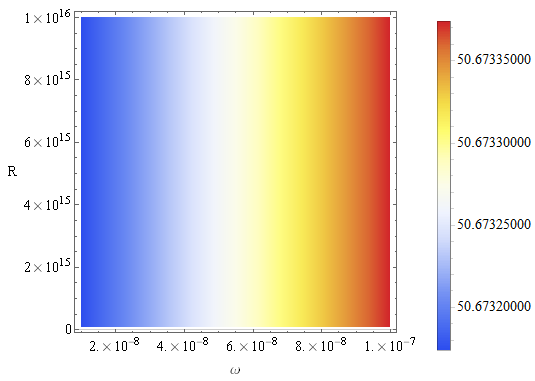}
		\caption{{\it{Shadow radius (left panel) and angular diameter (right panel) of the Sgr A$^\star$ black hole using the Model 1. We have set the location of the observer $r_0=10^{10}$ and $M_{0}=4.1 \times 10^6 M_{\odot}$.
 }}}
\label{Densityplot}
	\end{figure*}

\begin{figure*}[!htb]
 	\includegraphics[width=8 cm]{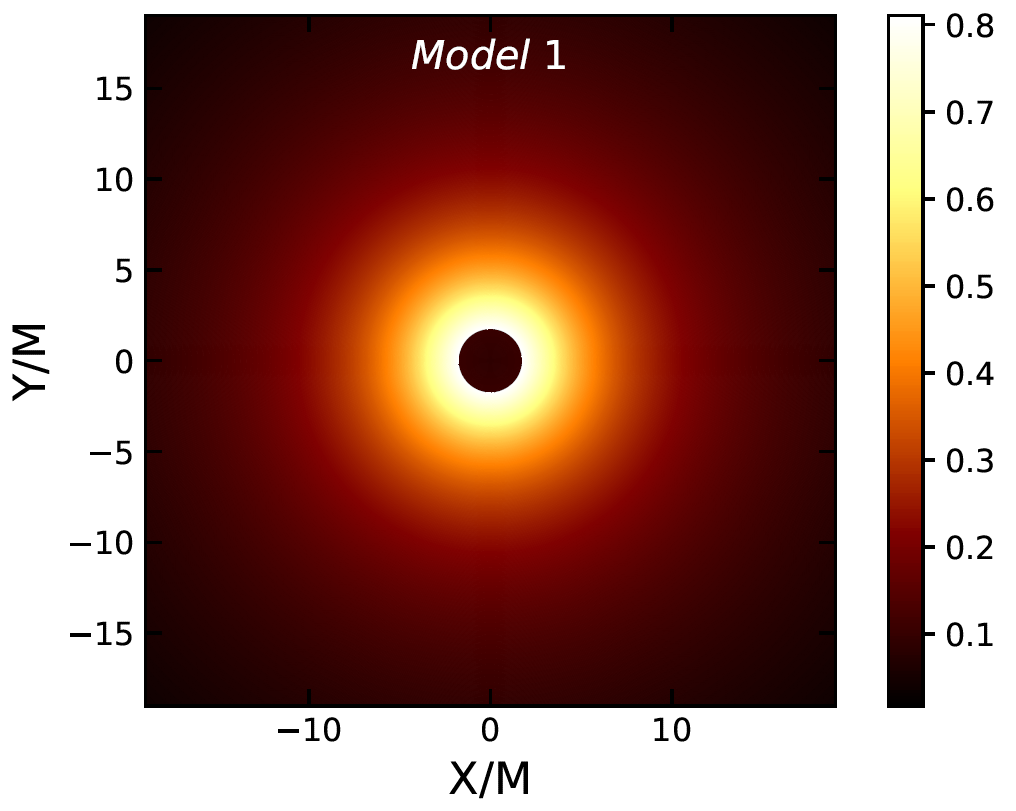}
 	\includegraphics[width=7.9 cm]{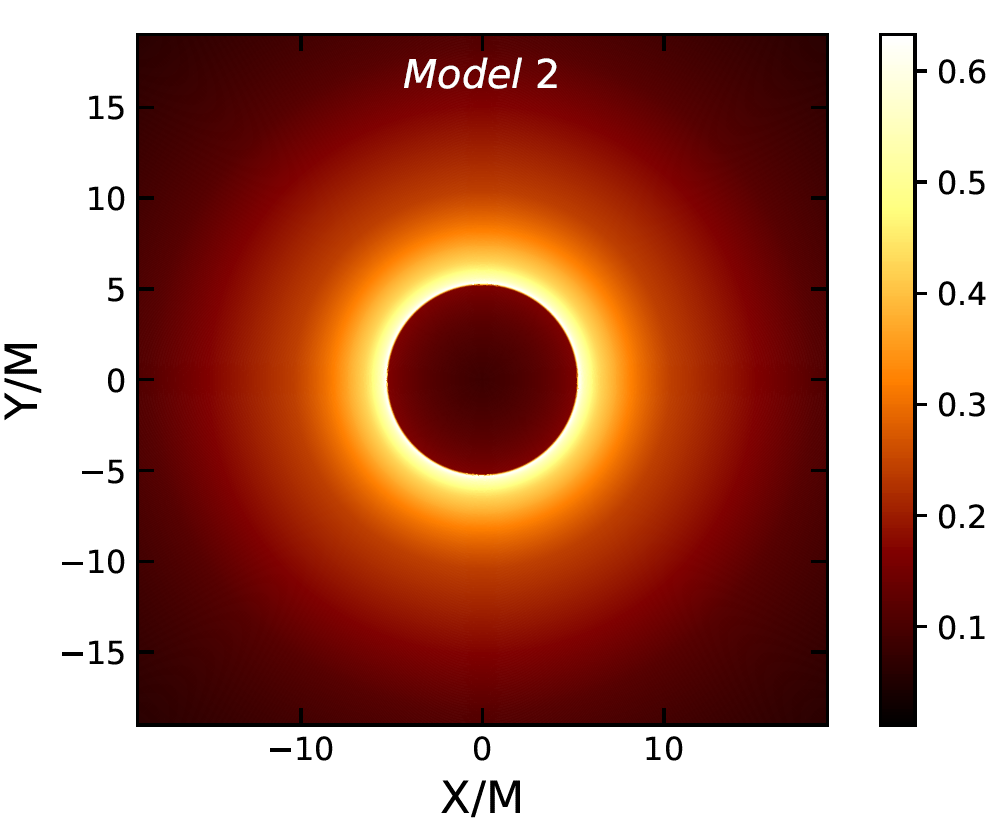}
		\caption{{Shadow images of the  Sgr A$^\star$ using the dark matter Model 1 and Model 2, respectively. We have used $\omega =2 \times 10^{-7}$, $M_{0}=4.1 \times 10^6 M_{\odot}$ and $R=10^{16} [M_{0}]$.}}
\label{Densityplot}
	\end{figure*}

\section{Dark matter effect on the shadow of black holes and naked singularites\label{secsha}}
We turn our focus now on the observational aspects of the our dark matter models with a black hole in the galactic center. Recently the studies concerning the shadow of black holes has recently gained a lot of interest \cite{sh1,sh2,sh3,sh4,sh5,sh6,sh7,sh8,sh9,sh10,sh11,sh12,sh13,sh14,sh15,sh16,sh17,sh18,sh19}. In the present work we would like to study the effect of the dark matter Model 1 and Model 2 on the shadow of black hole and naked singularity. We can start by writing the Hamilton-Jacobi equation
\begin{equation}
\frac{\partial S}{\partial \sigma}+H=0,
\end{equation}
with $S $ representing the Jacobi action and  $\sigma $  is some affine parameter along the geodesics. Considering the fact that our geometry is spherically symmetrical the Hamiltonian can be further written as
\begin{equation}
\frac{1}{2}\left[-\frac{p_{t}^{2}}{f(r)}+g(r)p_{r}^{2}+\frac{p_{\phi}^{2}}{r^{2}}\right] =0.
\label{EqNHa}
\end{equation}
	
To obtain the photon radius, first we use the fact that there are two constants of motions for the photon: the energy $E$ and angular momentum $L$, respectively. Secondly, to determine the circular and unstable orbits we can use the conditions given by the effective potentials
\begin{equation}
V_{\rm eff}(r) \big \vert_{r=r_{ph}}=0,  \qquad \frac{\partial V_{\rm eff}(r)}{\partial r}%
\Big\vert_{r=r_{ph}}=0,  
\end{equation}
It is not very difficult to show that \cite{sh18} 
\begin{equation}
\frac{dr}{d\phi}=\pm r\sqrt{g(r)\left[\frac{r^{2}}{b^2 f(r)} -1\right] },
\end{equation}
where $b$ is the impact parameter. Let us assume that a light ray is sent from a static observer located at some distance $r_{0}$ from the black hole, and let $\vartheta$ be the angle with respect to the radial direction, in that cas we can write \cite{sh18}
\begin{equation}
\cot \vartheta =\frac{\sqrt{g_{rr}}}{\sqrt{g_{\phi\phi}}}\frac{dr}{d\phi}\Big\vert%
_{r=r_{0}}.  \label{Eqangle}
\end{equation}
Without going into details here, one can show that the shadow radius measured by such an observer is computed via
\begin{equation}
R_{\rm sh}=r_{\rm ph}\sqrt{\frac{f(r_{0})}{f(r_{\rm ph})}}.
\end{equation}
	
On the other hand, from the condition of the unstable photon orbit$r_{\rm ph}$, for the Model 1 we get
\begin{eqnarray}
r^{\rm Model 1}_{\rm ph }=\frac{\left(4 \omega+1\right)M_0}{1-2 \omega},
\end{eqnarray}
and for the dark matter Model 2 we get
\begin{equation}
r^{\rm Model 2}_{\rm ph }=\frac{(12\omega-3\pm\sqrt{-48 \omega^2+24\omega+9})M_{BH}}{4 \omega-2}.
\end{equation}

One can us the photon orbit equations to compute the shadow radius and the angular diameter of the black hole and naked singularity. In Fig. 2 and 3, we have used both dark matter models and plotted the dependence of the shadow radius and angular diameter of the Sgr A$^{\star}$ on the state parameter $\omega$ and $R$. The parameter range used here is consistent with the constant flat curve. For the black hole (Model 2) we obtain larger shadow radius compared to the naked singularity (Model 1) which basically can distinguish these objects. However, the results show that the dark matter in the Model 2 increases the angular diameter of the black hole by the order of $10^{-4}\mu$arcsec compared to the Schwarzschild vacuum solution.  Such an effect is very small and out of reach of the  present technology. Next, we model the accretion disk as optically thick medium, with the specific intensity measured at some point $(X,Y)$ of the observer's image given by \cite{sh1,sh4,sh5,sh14,sh15,sh16,sh17}
\begin{eqnarray}
I_{obs}(\nu_{obs},X,Y) = \int_{\gamma}\mathrm{g}^3 j(\nu_{e})dl_\text{prop}.  
\end{eqnarray}
We also assume a rotating and radiating gas around the black hole with the four-velocity components 
\begin{eqnarray}
u^\mu_e=u^{t}\Big(1,0,0,\Omega \Big),
\end{eqnarray}
where $u^{t}=(f(r)-r^2 \Omega^2)^{-1/2}$, and $\Omega=\sqrt{f'(r)/2r}$. Moreover one has to use the relation for the photon $p_{\mu}p^{\mu}=0$, and the redshift function $\mathrm{g}$ given by
\begin{eqnarray}
  \mathrm{g} =\frac{p_{\mu}u_{obs}^{\mu}}{p_{\nu}u_e^{\nu}},
\end{eqnarray}
here $u_{obs}^{\mu}$ gives  the 4-velocity of the 
observer. In the present work, we assume that the specific emissivity is described by the radial law $r^{-2}$, according to
\begin{eqnarray}
    j(\nu_{e}) \propto \frac{\delta(\nu_{e}-\nu_{\star})}{r^2},
\end{eqnarray}
where $\nu_{\star}$ is the emitter's-rest frame frequency. We can now use the proper length to express the
total observed flux in terms of the radial coordinate \cite{sh1,sh4,sh5,sh14,sh15,sh16,sh17}
\begin{eqnarray} \label{inten}
    F_{obs}(X,Y) \propto -\int_{\gamma} \frac{\mathrm{g}^3 p_t}{r^2p^r}dr.  
\end{eqnarray}

In Fig. 4, we present the shadows and the corresponding intensities for the Model 1 and  Model 2, respectively. We see that for the black hole the size of the shadow is significantly larger. We can use the EHT result for the shadow radius of the Sgr A$^{\star}$ \cite{EHT2022-1,EHT2022-2,EHT2022-3,Vagnozzi:2022moj}
\begin{eqnarray}
4.54 \leq R_{sh}/M_0 \leq 5.22,
\end{eqnarray}
within $1\sigma $ confidence level, as well as 
\begin{eqnarray}
4.20 \leq R_{sh}/M_0 \leq 5.56
\end{eqnarray}
within $2\sigma $ confidence level.  We fix $\omega \simeq 2 \times 10^{-7}$ which is the tightest constraint obtained from the flat curves in Section III. Furthermore, we can fix the location of the observer at $r_0 \sim 10^{10}$ ( measured in units of mass $M_0$), then for Model 1 we obtain $R_{sh}/M_0\simeq 1.73$, while for the Model 2 we get $R_{sh}/M_0\simeq 5.196$, respectively. This simply means that the naked singularity (Model 1) is not consistent with observations and can be ruled out. The black hole model is in perfect agreement with the data, and the effect of dark matter are very small. Of course, we can consider $\omega$ as a free parameter and we can further use Eq. (61) to constrain $\omega$, say within $2 \sigma$ to obtain $\omega \leq 10^{-3}$. However, such constraint is not very precises and, as we will argue in the next section, such a bound cannot describe the experimental data for the S2 star orbit in our galaxy (leads to instability of the S2 orbit). \\

\section{Dark matter effect on the S2 star orbit}	
In this final section we would like to understand more about the stability of the S2 star orbit around the Sgr A black hole/naked singularity when the dark matter effect are taken into consideration.  Toward this goal, we can model the motion of the  S2 star as a point particle and, for simplicity, we are going to restrict our analyses in the equatorial plane $(\theta=\pi/2,~\dot\theta=0)$. From  the Lagrangian it follows that 
\begin{eqnarray}
		2\mathcal{L} &=& - f(r)\dot{t}^2
		+\frac{\dot{r}^2}{g(r) }+ r^2 \dot{\phi}^2.
\end{eqnarray}
In the present paper, we shall assume that the Sgr A$^\star$ black hole mass is $4.1 \times 10^6 M_{\odot}$ and will take $R \sim 10^{16}$. Next, we can use the above Lagrangian which implies two constants of motion: the energy of the particle $E$ and total angular 
momentum $L$, given by $E$ and $L$, respectively. Using the above results, it is not difficult to obtain the equation of motion for the S2 star  
\begin{eqnarray}\notag
\dot{t}&=&\frac{E}{f(r)},\\\notag
     \ddot{r} &=&\dfrac{1}{2} \left(g_{rr}(r)\right)^{-1}\left[\frac{d g_{tt}(r)}{dr} \dot{t}^2 +  
\frac{d g_{rr}(r)}{dr} \dot{r}^2 + \frac{d g_{\phi \phi}(r)}{dr} \dot{\phi}^2\right],\\
\dot{\phi}&=&\frac{L}{r^2}.\label{eqn:motionr}
\end{eqnarray}
	\begin{figure*}[!htb]
 	\includegraphics[width=5.1 cm]{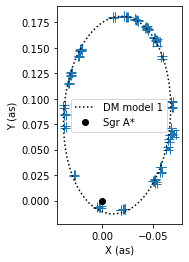}
 		\includegraphics[width=7.9 cm]{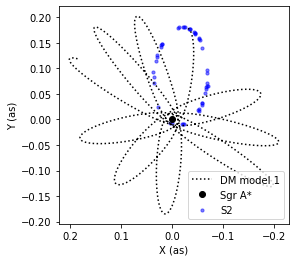}
		\caption{{Left panel: The stable orbit of S2 star having a naked singularity in the center using dark matter Model 1 with $\omega = 10^{-8}$, $M_{0}=4.1 \times 10^6 M_{\odot}$ and $R=10^{16} [M_{0}]$. Right panel: The unstable obit of S2 star using dark matter Model 1 for $\omega= 10^{-3}$, $M_{0}=4.10 \times 10^6 M_{\odot}$ and $R=10^{16} [M_{0}]$. In both cases we have used $\upsilon=66.26$, $i= 134.56$, and $\zeta=228.17$.
 }}
\label{Densityplot}
	\end{figure*}
	
	\begin{figure*}[!htb]
 	\includegraphics[width=5.1 cm]{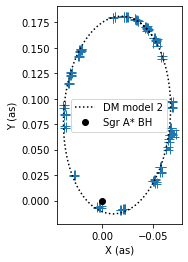}
 		\includegraphics[width=7.9 cm]{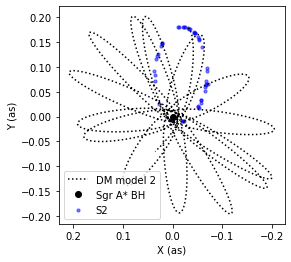}
		\caption{{Left panel: The stable orbit of S2 star having a black hole in the center using dark matter Model 2 with $\omega =2 \times 10^{-8}$, $M_{0}=4.1 \times 10^6 M_{\odot}$ and $R=10^{16} [M_{0}]$. Right panel: The unstable obit of S2 star using dark matter Model 1 for $\omega=2 \times 10^{-3}$, $M_{0}=4.10 \times 10^6 M_{\odot}$ and $R=10^{16} [M_{0}]$. In both cases we have used $\upsilon=66.26$, $i= 134.56$, and $\zeta=228.17$.
 }}
\label{Densityplot}
	\end{figure*}
We can use Cartesian coordinates to describe the motion of the S2 star in real orbit via $(x, y, z)$, along with the velocity components $(v_x, v_y , v_z)$, respectively. However, one can also use the coordinate transformation to relate the spherical coordinates to Cartesian coordinates by \cite{s1,s2,s3,s4,s5,s6,s7,s8}
\begin{align}\notag
   \{x,y,z\}&=\{r \cos\phi ,r \sin\phi,0\},\\\notag
    \{v_x,v_y\} &= \{v_r \cos\phi - r v_\phi \sin\phi , v_r \sin\phi + r v_\phi \cos\phi\}.
\end{align}
Furthermore, we need to define a new set of coordinates to describe the apparent orbit on the plane of the sky via  $(\mathcal{X}, \mathcal{Y}, \mathcal{Z})$ along with the apparent coordinate velocities using the relations \cite{s1,s2,s3,s4,s5,s6,s7,s8}
\begin{align}\notag
   \{\mathcal{X},\mathcal{Y}, \mathcal{Z} \}&= \{x B + y G,x A+ y F,x C + y F\}.\\\notag
    \{\mathcal{V}_X, \mathcal{V}_Y, \mathcal{V}_Z \}&=\{v_x B + v_y G,v_x A+ v_y F,v_x C + v_y F\},
\end{align}
where \cite{s1,s2,s3,s4,s5,s6,s7,s8}
\begin{eqnarray}\notag
    B &=& \sin\zeta \cos\upsilon + \cos\zeta \sin\upsilon \cos i \\\notag
    G &=& -\sin\zeta \sin\upsilon + \cos\zeta \cos\upsilon \cos i \\\notag
    A&=& \cos\zeta \cos\upsilon - \sin\zeta \sin\upsilon \cos i\\\notag
    F &=& -\cos\zeta \sin\upsilon - \sin\zeta \cos\upsilon \cos i\\\notag
    C&=& \sin\upsilon \sin i\\\notag
    F &=& \cos\upsilon \sin i.
\end{eqnarray}
In the above equations we have the following osculating orbital elements: the argument of pericenter ($\upsilon$), the inclination between the real orbit and the observation plane ($i$), and the  ascending node angle ($\zeta$), respectively. Finally, we can now use the spacetime geometry described by our dark matter model to find the orbit of the S2 star and compare it with the observational data. Using numerical methods (see for details \cite{s1,s2,s3,s4,s5,s6,s7,s8}), in Fig. 5 (left panel), e have found the elliptic orbit of the S2 star using our dark matter Model 1. For the dark matter parameter, we have used a specific value which is agreement with the constant flat curve for the velocity (as we found in Sec. 3). However, we find that by increasing the state parameter of the dark matter, more specifically when $\omega > 10^{-7}$, there is a significant effect on the orbit of the S2 star and the orbit is not stable to describe or confront the observational data for the S2 star. As a particular example, in Fig. 5 (right panel), we have shown the orbit of such a case using $\omega\sim 10^{-3}$. The results obtained from the Model 2, having a black hole in the center,  are very similar as can be seen from Fig 6.  
	
\section{Conclusions}
 In this paper we used the Einstein construction to model a black hole and a naked singularity immersed in a dark matter fluid with the parameter of state $\omega$. We studied two models which are spherically symmetric where the dark matter mass locally diverges but globally reduces to the asymptotically flat geometry. In order to achieve the asymptotically flat geometry we have matched the interior spacetime of the galaxy with the exterior metric models as Schwarzschild geometry and found that a thin shell of matter is needed.  As a special case, we have shown that for large distances from the galactic center the constant flat curves are obtained implying an upper bound for the dark matter state parameter $\omega \lesssim 10^{-7}$. We have also examined the the energy conditions for the dark matte fluid and showed that they are satisfied outside the black hole/naked singularity.  We used a rotating and radiating particles to model the optically thick medium and compute the shadow images.  Using astrophysical values for the dark matter parameter having a black hole in the galactic center, it is shown that the effect of dark matter fluid on the shadow radius is small, in particular, the angular radius of the black hole is shown to increase by the order $10^{-4} \mu$arcsec compared to the Schwarzschild vacuum solution. The black hole shadow (Model 2) is significantly larger compared to the naked singularity (Model 1) and can be distinguished with the present technology. Finally, we have shown that the effect of dark matter is very important on the motion of S2 star orbit, in particular the orbit is stable and compatible with the data for values within the range $\omega \lesssim 10^{-7}$, but further increase of $\omega $ leads to unstable orbits. From the EHT shadow images for the Sgr A$^{\star}$ black hole we obtained the following constrain $\omega \lesssim 10^{-3}$.  We used the tightest constraint obtained from the flat curves $\omega \simeq 2 \times 10^{-7}$ and found that the naked singularity (Model 1) is not consistent with the EHT  result. On the other hand, the  predicted shadow radius using the black hole solution with dark matter (Model 2) is in perfect agreement with observations. Let us point out here that, the state parameter $\omega$ can depend on the nature of the galaxy, namely different galaxies can have different state parameters. It is also possible that dark matter has different phases within a given galaxy, in that case we expect a richer dark matter phenomenology. In the near future, we expect more precise observations for the angular size and hence more precise constraints.

 \section*{Acknowledgments} I would like to thank Grant Remmen for discussions during the preparation of this manuscript.

\end{document}